\documentclass[%
preprint,
 10pt,
 nofootinbib,
 amsmath,
 amssymb,
 aps,
 prc,
]{revtex4-1}

\usepackage{graphicx}
\usepackage{dcolumn}
\usepackage{bm}
\usepackage{amsmath}

\usepackage[normalem]{ulem}
\usepackage{tikz}
\usetikzlibrary{decorations.pathmorphing}
\usetikzlibrary{decorations.markings}
\renewcommand\sout{\bgroup \color{red} \ULdepth=-.5ex \ULset}

\def\bs #1{{\boldsymbol #1}}

\begin{document}

\title{
New Universality for Near-Threshold Three-Body Resonances
}
\author{
Atsunari Konishi \\
{\it KEK Theory Center, IPNS, \\
High Energy Accelerator Research Organization (KEK), \\
1-1 Oho, Tsukuba, Ibaraki, 205-0801, Japan} \\
\vspace{0.5cm}
Osamu Morimatsu \\
{\it KEK Theory Center, IPNS, \\
High Energy Accelerator Research Organization (KEK), \\
1-1 Oho, Tsukuba, Ibaraki, 205-0801, Japan \\
Department of Physics, Faculty of Science, University of Tokyo, \\
7-3-1 Hongo Bunkyo-ku Tokyo 113-0033, Japan and \\
Department of Particle and Nuclear Studies, \\
Graduate University for Advanced Studies (SOKENDAI), \\
1-1 Oho, Tsukuba, Ibaraki 305-0801,Japan
} \\
\vspace{0.5cm}
Shigehiro Yasui \\
{\it Department of Physics, Tokyo Institute of Technology, \\
2-12-1 Ohokayama, Meguro, Tokyo, 152-8551, Japan \\
}
\vspace{0.5cm}
}
\affiliation{}
\date{\today}

\begin{abstract}
In the three-body system with one resonantly interacting pair, we study the behavior of the $S$-matrix pole near the threshold in the fourth quadrant of the unphysical complex energy plane.
Our study is essentially based on the unitarity and analyticity of the $S$-matrix
and employs the Alt-Grassberger-Sandhas (AGS) equations specifically for the three-body scattering problem and the dispersion relation for the inverse $T$-matrix.
We find that the trajectory of the complex energy, $E$, of the $S$-matrix pole near the threshold is uniquely given by
$c_0 + E \log{\left( - E \right)} \approx 0$
or
$c_0 + E_R \log E_R \approx 0$, $E_I \approx \pi E_R/\log E_R$
in the fourth quadrant of the unphysical complex energy plane,
in contrast to the non-unique trajectories with no resonantly interacting pair,
$c_0 + c_1 E + E^2 \log{\left( - E \right)} \approx 0$
or
$E_R \approx -c_0/c_1$, $E_I \approx -\pi E_R^2/c_1$
where $E_R$ and $E_I$ are the real and imaginary parts of $E$, respectively, and $c_0$ and $c_1$ are real constants.
This is a new universal behavior of the $S$-matrix near the threshold.
Also, we briefly discuss implications to exotic hadron candidates. 
\end{abstract}

\maketitle

It is known that the three-body system shows a remarkable behavior, the Efimov effect \cite{Efimov:1970zz, Efimov:1971zz, Braaten:2004rn,Naidon:2016dpf}, when two or three pairs interact resonantly, which means that the pair has a zero-energy bound state if isolated.
The Efimov effect means that in the three-body system an effective long-range three-body attraction is induced which generates infinitely many bound states with arbitrary small binding energies.
It is also known that the Efimov effect does not occur in the three-body system if only one pair interacts resonantly \cite{Amado:1992vn, efimov1972level, efimov1973energy}.
Then, one might ask if there is any characteristic behavior, precursor phenomena of the Efimov effect, in the three-body system with one resonantly interacting pair, which does not exist in the three-body system with no resonantly interacting pair.
With this question as a theoretical motivation, in the present paper we study the behavior of the $S$-matrix pole near the threshold in the three-body system with one resonantly interacting pair.
Our motivation comes from not only a purely theoretical one explained above but also a phenomenological one below.
Recently, many candidates of exotic hadrons have been experimentally observed \cite{Brambilla:2010cs}.
Exotic hadrons are hadrons which cannot be understood either as simple $q {\bar q}$ mesons or $qqq$ baryons.
We notice that some of exotic hadron candidates have masses which are close to both two-body and three-body thresholds of hadrons. 
If one pair interacts resonantly in the three-body system, two-body and three-body thresholds become degenerate.
Therefore, we hope to get some hints in order to understand such exotic hadron candidates by studying the $S$-matrix pole behavior near the threshold in the three-body system with one resonantly interacting pair.
This is our phenomenological motivation.

Our study is essentially based on the unitarity and analyticity of the $S$-matrix.
Specifically, we employ the Alt-Grassberger-Sandhas (AGS) equations \cite{Alt:1967fx} for the three-body scattering problem and the dispersion relation \cite{Bjorken:100770} for the inverse $T$-matrix.

We would also like to mention that in a separate paper \cite{Konishi:2017lbg} we introduce two-body and three-body channels as independent degrees of freedom
assuming their thresholds to be degenerate and study the behavior of the $S$-matrix pole near the threshold by numerically solving coupled-channels.

\subsection{Two-Body Sysytem}

Let us first review the partial wave $S$-matrix pole behavior near the threshold in the two-body system \cite{MR1947260}.
We consider the scattering of two different particles and ignore spins for simplicity.
From the unitarity of the $S$-matrix, the optical theorem for the partial wave $T$-matrix with angular momentum $\ell$, $t_{(\ell)} \left( E \right)$, is derived as
\begin{align}
	{\rm Im} \, t_{(\ell)} \left( E \right) &= \frac{t_{(\ell)} \left( E \right) - t_{(\ell)}^* \left( E \right)}{2i} 
	= t_{(\ell)}^* \left( E \right) \rho^{(2)} \left( E \right) t_{(\ell)} \left( E \right),
\label{1}
\end{align}
where $\rho^{(2)} \left( E \right)$ is the two-body density of states
\begin{align}
	\rho^{(2)} \left( E \right) & = \int d{\bs k}_1 d{\bs k}_2 \delta^3({\bs k}_1 + {\bs k}_2)\delta(E_1+E_2-E) \sim \sqrt{E}\Theta \left( E \right),
\label{2}
\end{align}
and $E$ is understood as the energy from the two-body threshold.
\footnote{Throughout this paper $E$ means the energy in the unit of the particle mass and $\approx$ means identical up to higher order terms while $\sim$ means identical up to higher order terms and an overall factor.}
Then, the imaginary part of the inverse partial wave $T$-matrix, $t_{(\ell)}^{-1} \left( E \right)$, is obtained by dividing Eq.~(\ref{1}) by $t_{(\ell)}^* \left( E \right)$ and $t_{(\ell)} \left( E \right)$ as
\begin{align}
	{\rm Im} \, t_{(\ell)}^{-1} \left( E \right) &= \frac{t_{(\ell)}^{-1} \left( E \right) - t_{(\ell)}^{*-1} \left( E \right)}{2i} 
	= -\rho^{(2)} \left( E \right).
\label{3}
\end{align}
Since  $t_{(\ell)} \left( E \right) \sim E^{\ell}$ we consider $E^{\ell}t_{(\ell)}^{-1} \left( E \right)$, not $t_{(\ell)}^{-1} \left( E \right)$, in order to discuss the partial wave $S$-matrix pole behavior near the threshold. 
As a complex function, the singular term of $E^{\ell}t_{(\ell)}^{-1} \left( E \right)$ is given by
\begin{align}
	\left[ E^{\ell}t_{(\ell)}^{-1} \left( E \right) \right]_{\rm sing.} = \frac{1}{\pi} \int^{\infty}_0 dE' \frac{{\rm Im} \, E'^{\ell}t_{(\ell)}^{-1} \left( E' \right)}{E' - E} \sim E^\ell\sqrt{-E}.
\label{4}
\end{align}
With regular terms included, $E^{\ell}T^{\ell-1} \left( E \right)$ behaves near the three-body threshold as
\begin{align}
	E^{\ell}t_{(\ell)}^{-1} \left( E \right) \sim c_0 + \cdots + c_\ell E^\ell + E^\ell\sqrt{-E},
\label{5}
\end{align}
where $O(E^{\ell+1})$ regular terms are neglected since they are of higher order than the singular term, $E^\ell\sqrt{-E}$.
From $E^{\ell}t_{(\ell)}^{-1} \left( E \right)=0$, the complex energy of the $S$-matrix pole near the three-body threshold satisfies
\begin{align}
	& c_0 + \cdots + c_\ell E^\ell + E^\ell\sqrt{-E} \approx 0,
\label{6}
\end{align}
or its real and imaginary parts, $E_R$ and $E_I$, satisfy
\begin{align}
	& E_R \approx - c_0^2,  \qquad E_I \approx 0 \qquad (\ell = 0), \\
	& E_R \approx -\frac{c_0}{c_1}, \qquad E_I \approx - \frac{1}{c_1} E_R^{\ell+\frac{1}{2}} \qquad (\ell \ge 1).
\end{align}
The behavior of the $S$-matrix pole in the $S$-wave, but not in higher partial waves, is universal in the sense that the $S$-matrix pole trajectory is uniquely determined independent of parameters, $c_0, \cdots, c_{\ell}$.


\begin{figure}[hp]
\begin{center}
\begin{tikzpicture}[domain=0:3.5, samples=100, >=stealth]
\draw (2.5,2.5) node[below left]{(a)}; 
\draw (0,0) node[below left]{threshold}; 
\draw (1.25,0) node[above]{cut}; 
\path (-3.5,0)--(3.5,0);
\draw [->] (-2.5,0)--(2.5,0) node[below] {$E_R$};
\path (0,-3.5)--(0,3.5);
\draw [->] (0,-2.5)--(0,2.5) node[left] {$E_I$};
\draw [decorate,decoration=zigzag] (0,0)--(2.5,0);
\draw [very thick] (0,0)--(-2.5,0) node [above] {$E_I=0$};
\end{tikzpicture}
\begin{tikzpicture}[domain=0:3.5, samples=100, >=stealth]
\draw (2.5,2.5) node[below left]{(b)}; 
\draw (0,0) node[below left]{threshold};
\draw (1.25,0) node[above]{cut}; 
\path (-3.5,0)--(3.5,0);
\draw[thin, ->] (-2.5,0)--(2.5,0) node[below] {$E_R$};
\path (0,-3.5)--(0,3.5);
\draw[->] (0,-2.5)--(0,2.5) node[left] {$E_I$};
\draw [decorate,decoration=zigzag] (0,0)--(2.5,0) ;
\draw [very thick,domain=0:2.5] plot(\x, {-0.6*sqrt(\x^3)});
\draw [very thick,domain=0:1.84202] plot(\x, {-sqrt(\x^3)}) node [below] {$\displaystyle{E_I=-\frac{1}{c_1}E_R^{\frac{3}{2}}}$};
\draw [very thick,domain=0:1.1604] plot(\x, {-2*sqrt(\x^3)});
\end{tikzpicture}
\caption{The $S$-matrix pole trajectories near the threshold in the unphysical sheet of the complex energy plane in the two-body system of (a) $s$-wave and (b) $p$-wave.
In the case (b) different trajectories correspond to different values of the parameter, $c_1$, while in the case (a) the trajectory is unique independent of the parameters.}
\end{center}
\end{figure}
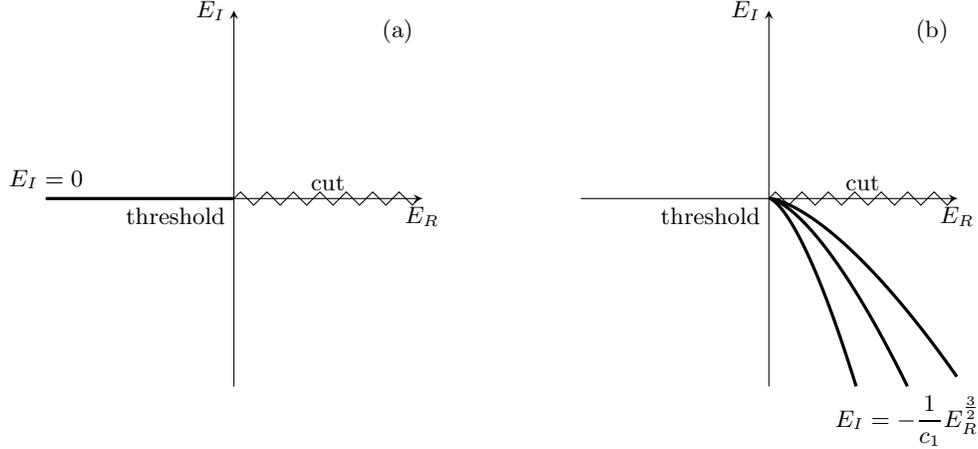

\subsection{Three-Body System}

We now move on to the three-body system with vanishing angular momentum.
The discussions proceed in a parallel manner as those in the two-body system.
We consider three different particles and ignore spins.
We assume that each pair has only one bound state just for notational simplicity.
We note two-body scattering states of a particle and a bound state of other two particles as $|1\rangle = |1(23)\rangle $, $|2\rangle  = |2(31)\rangle $, $|3\rangle  = |3(12)\rangle $ and the three-body scattering state as $|0\rangle $.

We start from the AGS equations \cite{Alt:1967fx} for the three-body scattering problem. 
The operator $U_{ij} \left( E \right)$ $(i,j=1,2,3)$ of the three-body scattering satisfies the AGS equations,
\begin{align}
	U_{ij} \left( E \right) = {\bar \delta}_{ij} G_0^{-1} \left( E \right) + \sum_{k = 1}^3 {\bar \delta}_{ik} t_k \left( E \right) G_0 \left( E \right) U_{kj} \left( E \right),
\end{align}
where ${\bar \delta}_{ij} = 1 - \delta_{ij}$, $G_0 \left( E \right)$ is the free three-body Green function, $t_k \left( E \right)$ is the two-body $T$-matrix of particles $i$ and $j$.
The imaginary part of the three-body scattering operator, $U_{ij} \left( E \right)$, is shown to satisfy \cite{Afnan:1977pi}
\begin{align}
	{\rm Im} \, U_{ij} \left( E \right) = \sum_{k = 1}^3 U_{ik}^{\dagger} \left( E \right) \rho^{(2)}_k \left( E \right) U_{kj} \left( E \right) + U_{i0}^{\dagger} \left( E \right) \rho^{(3)} \left( E \right) U_{0j} \left( E \right),
\label{10}
\end{align}
where
\begin{align}
	U_{0i} \left( E \right) = \sum_{k = 1}^3 t_k \left( E \right) G_0 \left( E \right) U_{ki} \left( E \right).
\end{align}
$\rho^{(2)}_k(E)$ and $\rho^{(3)}(E)$ are the two-body and three-body density of states, respectively,
\begin{align}
	\rho^{(2)}_k \left( E \right) & = \int d{\bs k}_k d{\bs k}_{ij} \delta^3({\bs k}_k + {\bs k}_{ij})\delta(E_k+E_{ij}-E) \sim \sqrt{E}\Theta \left( E \right),
	\label{12} \\
	\rho^{(3)}  \left( E \right) & = \int d{\bs k}_i d{\bs k}_j d{\bs k}_k  \delta^3({\bs k}_i + {\bs k}_j + {\bs k}_k) \delta(E_i+E_j + E_k - E) \sim E^2 \Theta\left( E \right),
	\label{13}
\end{align}
where ${\bs k}_k$ and $E_k$ (${\bs k}_{ij}$ and $E_{ij}$) are the momentum and the energy of the particle $k$ (the bound state of particles $i$ and $j$).
$E$ is understood in Eq.~(\ref{12}) as the energy from the threshold of a particle and a bound state of other two particles and in Eq.~(\ref{13}) as the energy from the three-body threshold.

$T$-matrices for a two-body to two-body process, $i + (jk) \rightarrow j + (ki)$, and a two-body to three-body process, $i + (jk) \rightarrow i + j + k$, are respectively given by the matrix elements, $T_{ji} \left( E \right) = \left<j| U_{ji} \left( E \right) |i\right>$ and $T_{0i} \left( E \right) = \left<0| U_{0i} \left( E \right) |i\right>$.

It should be noted that Eq.~(\ref{10}) can be cast into a more familiar form of the optical theorem for the forward elastic scattering amplitude of $i + (jk) \rightarrow i + (jk)$ as \cite{Amado:1967zz}
\begin{align}
	{\rm Im} \, T_{ii} \left( E \right) = \sum_{k = 1}^3 T_{ik}^{\dagger} \left( E \right) \rho^{(2)}_{k} \left( E \right) T_{ki} \left( E \right) + T_{i0}^{\dagger} \left( E \right) \rho^{(3)} \left( E \right) T_{0i} \left( E \right).
\label{14}
\end{align}

Then, the imaginary part of the inverse three-body scattering operator, $U^{-1}_{ij} \left( E \right)$, is given by
\begin{align}
	{\rm Im} \, U^{-1}_{ij} \left( E \right) 
	&= -\frac{1}{2} \rho^{(3)} \left( E \right) + \delta_{ij} \left\{ \sum_k \rho^{(2)}_{k} \left( E \right) + \left[1 + t_j^{\dagger} \left( E \right) G_0^{\dagger} \left( E \right) \right] \rho^{(3)} \left( E \right) \Big[ G_0 \left( E \right) t_j \left( E \right) + 1 \Big] \right\}, 
\label{15}
\end{align}
from which the singular part of $U^{-1}_{ij} \left( E \right)$ is determined through the dispersion relation as
\begin{align}
	U^{-1}_{ij} \left( E \right) = \frac{1}{\pi} \int dE' \frac{{\rm Im} \, U^{-1}_{ij} \left( E \right)}{E' - E}.
\label{16}
\end{align}
Having identified the singular part of $U^{-1}_{ij} \left( E \right)$, with the regular terms of  $U^{-1}_{ij} \left( E \right)$ the $S$-matrix pole energy is obtained as a solution of the equation,
\begin{align}
	{\rm det} \, U^{-1} \left( E \right) = 0.
\label{17}
\end{align}

\subsubsection{The case with no resonantly interacting pair}

First, we consider the case in which no pair in the three-body system interacts resonantly, namely the three-body and two-body channels are not degenerate.
Since two-body $T$-matrices $t_k \left( E \right)$ are regular at the threshold,
the imaginary part of $U^{-1}_{ij} \left( E \right)$ near the three-body threshold is entirely determined by the three-body density of states $\rho^{(3)} \left( E \right)$ as
\begin{align}
	{\rm Im} \, U^{-1}_{ij} \left( E \right) \sim \rho^{(3)} \left( E \right) \sim E^2 \Theta \left( E \right).
\label{18}
\end{align}
Therefore, as a complex function the singular term of $U_{ij}^{-1} \left( E \right)$ becomes
\begin{align}
	\left[ U^{-1}_{ij} \left( E \right) \right]_{\rm sing.} \sim \frac{1}{\pi} \int^{\infty}_0 dE' \frac{E'^2}{E' - E} \sim E^2 \log{\left( - E \right)}.
\label{19}
\end{align}
With regular terms included, $\det U^{-1} \left( E \right)$ behaves near the three-body threshold as
\begin{align}
	\det U^{-1} \left( E \right) \sim c_0 + c_1 E + E^2 \log{\left( - E \right)},
\label{20}
\end{align}
where $O(E^2)$ regular terms are neglected since they are of higher order than the singular term, $E^2 \log{\left( - E \right)}$.

From $\det U^{-1} \left( E \right)=0$, the complex energy of the  $S$-matrix near the three-body threshold satisfies
\begin{align}
	c_0 + c_1 E + E^2 \log{\left( - E \right)} \approx 0,
\label{21}
\end{align}
or its real and imaginary parts, $E_R$ and $E_I$, satisfy
\begin{eqnarray}
	&& E_R \approx -\frac{c_0}{c_1}, \qquad E_I \approx -\frac{\pi}{c_1} E_R^2,
\label{22}
\end{eqnarray}
in the fourth quadrant of the unphysical complex energy plane,
which was shown in Ref. \cite{Matsuyama:1991bm}.

\subsubsection{The case with one resonantly interacting pair}
Secondly, we consider the case in which a pair in the three-body system interacts resonantly, namely the three-body and two-body channels are degenerate.
Since the two-body $T$-matrix of the pair, which has a zero-energy bound state, is singular at low energy as can be seen from Eq.~(\ref{5}) as
\begin{align}
	t_k \left( E \right) \sim \frac{1}{\sqrt{E}}.
\label{23}
\end{align}
The leading behavior of ${\rm Im} \, U_{ij}^{-1} \left( E \right)$ therefore comes from the second term in Eq.~(\ref{15}), specifically from the term $G_0^{\dagger} \left( E \right) t_k^{\dagger} \left( E \right) \rho_2 \left( E \right)  t_k \left( E \right) G_0 \left( E \right)$ as
\begin{align}
	{\rm Im} \, U_{ij}^{-1} \left( E \right) & \sim t_k^{\dagger} \left( E \right) \rho_2 \left( E \right) t_k \left( E \right)
	\sim \frac{1}{\sqrt{E}} E^2 \Theta \left( E \right) \frac{1}{\sqrt{E}}
	\sim E \Theta \left( E \right).
\label{24}
\end{align}
Then, as a complex function the leading singular term of $U^{-1}_{ij} \left( E \right)$ becomes
\begin{align}
	\left[ U^{-1}_{ij} \left( E \right) \right]_{\rm sing.} \sim \frac{1}{\pi} \int^{\infty}_0 dE' \frac{E'}{E' - E} \sim E \log{\left( - E \right)}.
\label{25}
\end{align}
With regular terms included, $\det U^{-1} \left( E \right)$ behaves near the three-body threshold as
\begin{align}
	\det U^{-1} \left( E \right) \sim c_0 + E \log{\left( - E \right)}.
\label{26}
\end{align}
Now, $O(E)$ regular terms are neglected since they are of higher order than the singular term, $E \log{\left( - E \right)}$.

From $\det U^{-1} \left( E \right)=0$, the complex energy of the  $S$-matrix near the three-body threshold satisfies
\begin{align}
	c_0 + E \log{\left( - E \right)} \approx 0,
\label{27}
\end{align}
or its real and imaginary parts, $E_R$ and $E_I$, satisfy
\begin{align}
	c_0 + E_R \log E_R \approx 0, \qquad E_I \approx \frac{\pi E_R}{\log E_R},
\label{28}
\end{align}
in the fourth quadrant of the unphysical complex energy plane.
Eq.~(\ref{27}) or Eq.~(\ref{28}) is the main result of this paper.


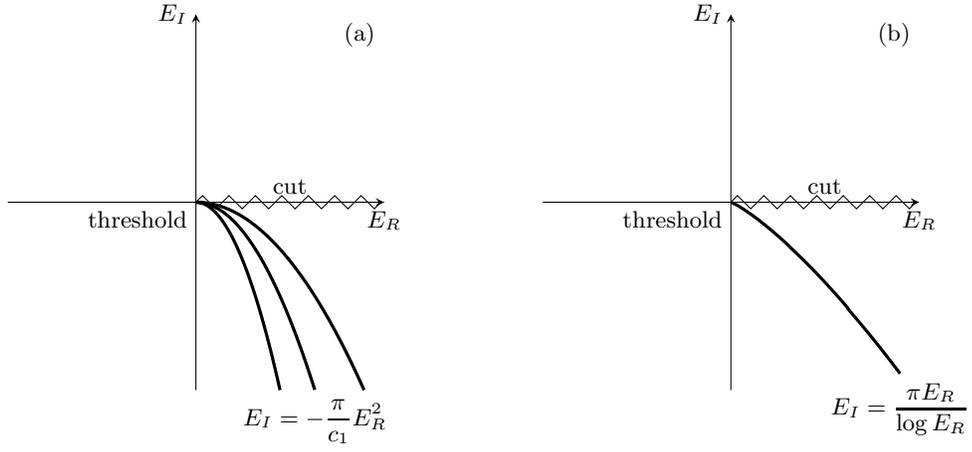
\begin{figure}[hp]
\begin{center}
\begin{tikzpicture}[domain=0:3.5, samples=100, >=stealth]
\draw (2.5,2.5) node[below left]{(a)}; 
\draw (0,0) node[below left]{threshold};
\draw (1.25,0) node[above]{cut};
\path (-3.5,0)--(3.5,0);
\draw[thin, ->] (-2.5,0)--(2.5,0) node[below] {$E_R$};
\path (0,-3.5)--(0,3.5);
\draw[->] (0,-2.5)--(0,2.5) node[left] {$E_I$};
\draw [decorate,decoration=zigzag] (0,0)--(2.5,0);
\draw [very thick,domain=0:2.23607] plot(\x, {-0.5*\x^2});
\draw [very thick,domain=0:1.58114] plot(\x, {-\x^2}) node [below] {$\displaystyle{E_I=-\frac{\pi}{c_1}E_R^2}$};
\draw [very thick,domain=0:1.11803] plot(\x, {-2*\x^2});
\end{tikzpicture}
\begin{tikzpicture}[domain=0:35, samples=100, >=stealth]
\draw (2.5,2.5) node[below left]{(b)}; 
\draw (0,0) node[below left]{threshold};
\draw (1.25,0) node[above]{cut};
\path (-3.5,0)--(3.5,0);
\draw [->] (-2.5,0)--(2.5,0) node[below] {$E_R$};
\path (0,-3.5)--(0,3.5);
\draw [->] (0,-2.5)--(0,2.5) node[left] {$E_I$};
\draw [decorate,decoration=zigzag] (0,0)--(2.5,0);
\draw [very thick,domain=0.001:2.25] plot(\x, {(pi*\x)/ln(\x/50)}) node[below] {$\displaystyle{E_I=\frac{\pi E_R}{\log E_R}}$};
\end{tikzpicture}
\caption{The $S$-matrix pole trajectories near the threshold in the unphysical sheet of the complex energy plane in the three-body system, when (a) no pair and (b) one pair interacts resonantly.
In the case (a) different trajectories correspond to different values of the parameter, $c_1$, while in the case (b) the trajectory is unique independent of the parameters.}
\end{center}
\end{figure}

It should be noted that when no pair interacts resonantly, case (a), different trajectories correspond to different values of the parameter, $c_1$, while when one pair interacts resonantly, case (b), the trajectory is unique independent of the parameters.
This behavior of the $S$-matrix pole in the three-body system is universal in the sense that it does not depend on details of the interactions except that one pair in the system is resonantly interacting.

As far as we know this is a new universal behavior of resonances near the threshold which has not been discussed before.
These resonances may have phenomenological impacts in the following sense.
If a two-body system is in the $S$-wave it cannot have resonances near the threshold since the $S$-matrix pole lies on the negative real axis in the unphysical complex energy sheet as seen in Fig.~1 (a).
But if the two-body system is coupled to the three-body system and their thresholds are degenerate, it corresponds to the case with one resonantly interacting pair, case (b).
Therefore, near-threshold resonances can exist since the $S$-matrix pole lies in the fourth quadrant of the unphysical complex energy sheet as seen in Fig.~2 (b).

It is not clear, however, if the universal behavior found in the present paper has something to do with experimentally observed exotic hadron candidates, in particular $X(3872)$ \cite{Choi:2003ue}.
It is partly because there exist angular momentum excitations in experimentally observed exotic hadron candidates, which are not considered in the present paper.
Therefore, it would be our future problem to study the near-threshold behavior of the $S$-matrix pole with including angular momentum excitations together with realistic applications to  experimentally observed exotic hadron candidates.

In conclusion, we have studied a three-body system in which one of three pairs interacts resonantly and found a new universal behavior of the $S$-matrix pole behavior near the threshold. The results we have obtained might play an important role in understanding excited hadrons,  which lie just above the degenerate threshold of two-body and three-body hadrons, especially candidates for the exotic hadrons.

We would like to thank Pascal Naidon for discussions about Efimov physics in general and also useful comments on our work.
We would also like to thank Hiroyuki Kamano, Toru Sato and Koichi Yazaki for helpful discussions.
\bibliography{ref.bib}

\end{document}